\documentclass[a4paper,11pt]{report}
\usepackage{graphicx}
\usepackage[numbers,compress]{natbib}
\usepackage{amssymb,amsmath,amsfonts}
\usepackage{hyperref}
\graphicspath{{./figures/}}
\title{}
\date{}

\begin{document}


\chapter[The circular jump...]{The circular jump as a hydrodynamic white hole}

{\Large G. Jannes}\footnote{jannes@ltl.tkk.fi}\\
\textit{Low Temperature Laboratory, Aalto University School of Science, PO Box 15100, 00076 Aalto, Finland\newline
Universit\'{e} de Nice Sophia Antipolis, Laboratoire J.-A. Dieudonn\'{e}, UMR
CNRS-UNS 6621, Parc Valrose, 06108 Nice Cedex 02, France}\vspace*{1ex}\\
{\Large G. Rousseaux}\footnote{Germain.Rousseaux@unice.fr}\\
\textit{Universit\'{e} de Nice Sophia Antipolis, Laboratoire J.-A. Dieudonn\'{e}, UMR
CNRS-UNS 6621, Parc Valrose, 06108 Nice Cedex 02, France}\\

\textbf{Abstract:} Surface waves in classical fluids experience a rich array of black/white hole horizon effects. The dispersion relation depends on the characteristics of the fluid as well as on the fluid depth and the wavelength regime. We focus on the shallow-water regime, and discuss the experimental proof that the circular hydraulic jump marks the transition between a supercritical and a subcritical flow regime. This finally confirms a theoretical conjecture formulated by Lord Rayleigh nearly 100 years ago. It also confirms that the circular jump corresponds to the spontaneous formation of a hydrodynamic white hole, with interesting characteristics from the point of view of analogue gravity. We study the dispersive regime, mention some lessons about the trans-Planckian issue and describe possible directions for future work.


\section{Introduction}
Surface waves in classical fluids provide a natural and rich class of black/white hole analogues. A familiar example of a hydrodynamic {\it black} hole analogue is the draining bathtub. From an experimental point of view, this has the obvious complication that the interesting physics literally disappears down the drain, and is unavoidably mixed with effects such as vorticity and turbulence which complicate the analysis. But many properties of black holes are directly related to those of white holes through a simple time-inversion, at least in the non-dispersive and non-dissipative limit. Experimental efforts have therefore so far concentrated mainly on {\it white} hole analogues. Two familiar examples are the blocking of sea waves at a river mouth and the approximately circular jump created by opening the tap in a kitchen sink. These two types of white hole analogues can be reproduced in controlled laboratory settings in order to study the associated horizon effects and their possible lessons for relativity (and vice versa: lessons from relativity for fluid mechanics). The river-mouth example corresponds to deep water waves propagating against a counter-current in a wave channel, while the kitchen-sink example corresponds to shallow waves on a circular hydraulic jump. 

The general dispersion relation for capillary-gravity surface waves propagating against a counter-current of velocity $U$ (see, e.g.,~\cite{Rousseaux:2010md} and references therein) is 
\begin{equation}\label{disp-rel}
 (\omega-{\bf U.k})^2=\left(gk+\frac{\gamma}{\rho}k^3\right)\tanh(kh),
\end{equation}
with $g$ the gravitational constant, $\gamma$ the surface tension, $\rho$ the density, $h$ the fluid depth, and as usual $\omega$ and $k$ are the frequency and wavenumber, respectively. From now on we will neglect any tangential effects and consider only surface waves propagating linearly (or radially) against the background flow.  The ``relativistic'' regime is obtained in the shallow-water limit ($kh\ll 1$) of gravity-wave propagation ($\gamma=0$): $(\omega -Uk)^2 \simeq ghk^2$, with the relativistic ``invariant'' speed $c=\sqrt{gh}$~\cite{Schutzhold:2002rf}.

It is also instructive to study deviations from this relativistic regime, where the propagation velocity of the surface waves is no longer invariant.  From now on, we will therefore refer to a ``horizon'' in the generalized sense of the point (or region) where the group velocity $c_g\equiv d\omega/dk$ of the surface waves vanishes. Indeed, such deviations can provide useful clues with regard to quantum gravity phenomenology and the robustness of black hole physics (for instance, Hawking radiation) with respect to possible non-relativistic features of the trans-Planckian physics. Besides Hawking radiation, other applications include the study of quasinormal modes (QNMs): the decay or energy dissipation modes of a perturbed black-hole spacetime. Such QNMs can be studied analytically and numerically, for example for surface waves (or sound waves) in the draining bathtub geometry~\cite{Berti:2004ju,Dolan:2011ti}. ``Trans-Planckian'' corrections to this QNM spectrum could then be similar to those obtained for acoustic black holes in Bose-Einstein condensates~\cite{jannes-QNMs}, since the high-wavenumber dispersion relation of surface waves has a superluminal $k^4$-dominance similar to the BEC dispersion relation 
\begin{equation}\label{disp-BEC}
 (\omega-Uk)^2=c_s^2\left(k^2+\frac{1}{4}\xi^2 k^4\right),	
\end{equation}
where $c_s$ is the speed of sound and $\xi \equiv \hbar/(mc_s)$ represents the healing length of the condensate.

We will now focus on a typical example of a shallow-water geometry, namely the circular hydraulic jump.

\section{The circular hydraulic jump}
\begin{figure}[!htpb]
 \centering
\includegraphics[width=.55\textwidth]{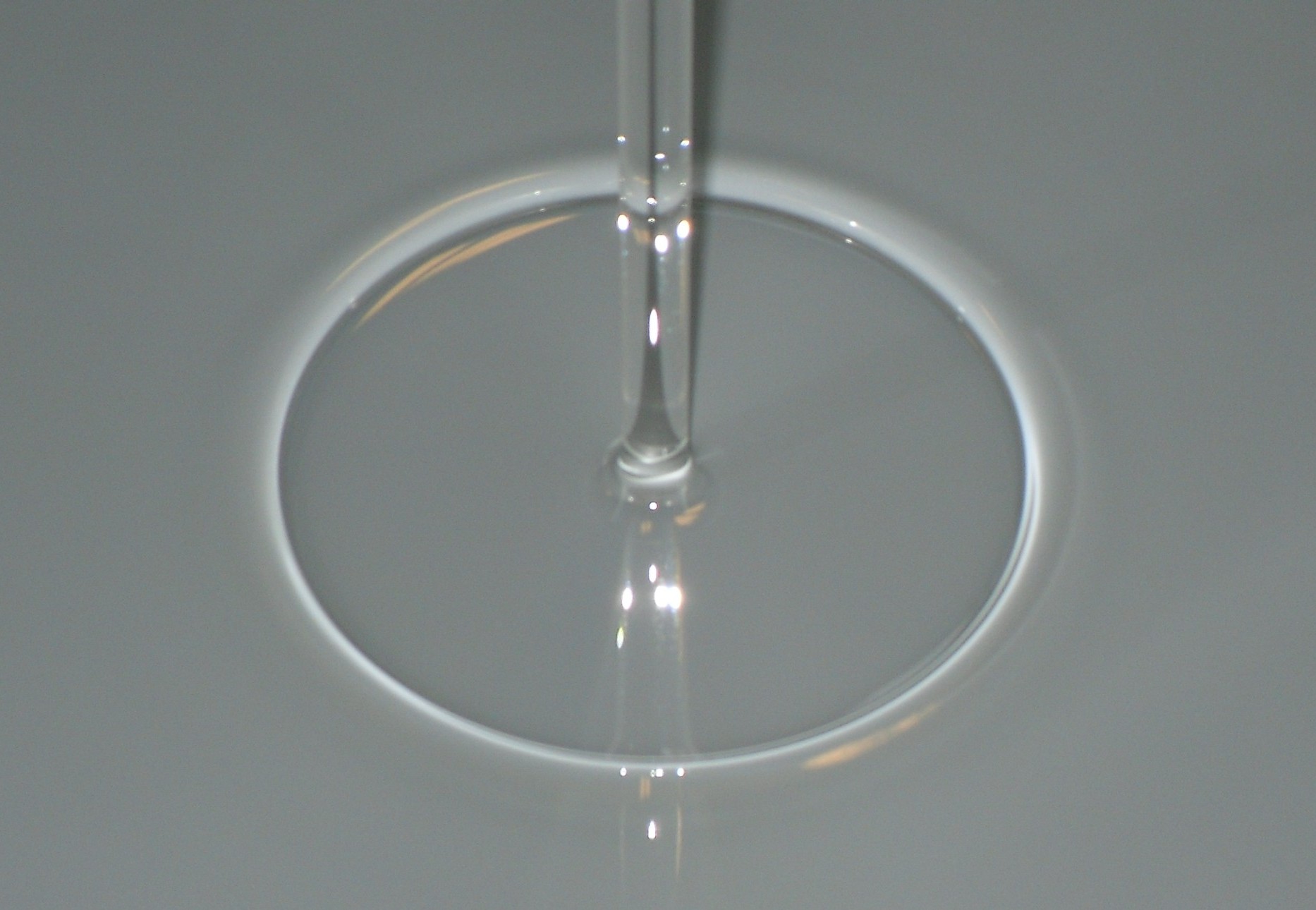}
 \caption{The circular hydraulic jump.}
 \label{Fig:jump}
\end{figure}

\begin{figure}[!htpb]
 \centering
\includegraphics[width=.75\textwidth]{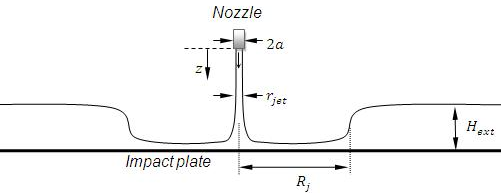}
 \caption{Basic setup of the circular hydraulic. A liquid is pumped through a nozzle and the fluid jet impacts vertically onto a horizontal plate. At a distance $R_j$, a sudden ``jump'' occurs in the fluid height.}
 \label{Fig:jump-schema}
\end{figure}

When a vertical fluid jet impacts on a horizontal plate with a sufficient flow rate, it will form a thin layer near the impact zone, which expands radially and at a certain distance abruptly increases in thickness, see Figs.~\ref{Fig:jump} and ~\ref{Fig:jump-schema}. This phenomenon is known as the circular hydraulic jump, and was already described by Leonardo Da Vinci. Its first modern description is due to Lord Rayleigh~\cite{Rayleigh:1914}, who developed a momentum-balance theory to describe it, and hypothesized that its appearance is due to a deceleration from a supercritical flow in the inner region to a subcritical flow outside. In other words, in the inner region, the radial fluid velocity at the surface $v_r^s$ (which plays the role of the counter-current $U$ in the dispersion relation~\eqref{disp-rel} above) is larger than the propagation speed of the surface waves $c$, so that these surface ripples can only propagate downstream. In the outer region, $v_r^s<c$ and hence the ripples can propagate in both directions. The transition would be characterised precisely by the jump, analogous to a shock wave. The circular jump is therefore the stationary wave solution or zero mode (zero-frequency solution of the dispersion relation) in the case of a circularly propagating, decelerating fluid flow.

The essential point in the context of analogue gravity is similar to the case of acoustic black holes~\cite{Unruh:1980cg,Visser:1997ux} and other examples of analogue gravity~\cite{Barcelo:2005fc}: the propagation of these surface waves obeys a generalised d'Alembertian equation in which the intervening curved-spacetime metric is the 2+1-di\-men\-sion\-al Painlev\'{e}-Gullstrand-Lema\^{i}tre (PGL) form of the Schwarzschild metric~\cite{Ray:2007}. Indeed, the line element for the circular jump is
\begin{equation}
ds^2=\frac{1}{c^2}\left[[c^2-(v_r^s)^2]dt^2+ 2v_r^s dt\, dr -dr^2-r^2d\phi^2\right],
\end{equation}
where the surface wave propagation speed $c$ plays the role of the speed of light in gravity, while the radial surface flow velocity $v_r^s$ corresponds to the local velocity of an observer freely falling from infinity in the case of gravity. The PGL metric
\begin{equation}
g_{\mu\nu}=
\begin{pmatrix}
c^2-(v_r^s)^2&& v_r^s\\v_r^s&&-I
\end{pmatrix}
\end{equation}
with $I$ the unit matrix, transforms into the Schwarzschild form through the coordinate transformation $d\tilde{t}=dt+dr\, v_r^s/[c^2-(v_r^s)^2]$. The condition $g_{tt}=0$ for a horizon in the non-rotating case becomes simply $c=v_r^s$. It is in this precise sense that the circular jump is believed to constitute the hydrodynamical analogue of a white hole \cite{Volovik:2005ga,Volovik:2006cz}.

Ever since Watson improved Rayleigh's account by incorporating the effects of viscosity~\cite{Watson:1964}, there has been a continuous experimental and theoretical interest in the circular jump, with a remarkable burst of work in the last decade. Curiously, though, an experimental proof of Rayleigh's original conjecture that the jump corresponds to a supercritical-to-subcritical transition (and hence, in the language of analogue gravity, that the circular jump constitutes a white hole) was only provided recently through the Mach cone experiment~\cite{Jannes:2010sa} that we describe next.

\section{Mach cone experiment}
\begin{figure}[!htpb]
 \centering
\includegraphics[width=.85\textwidth]{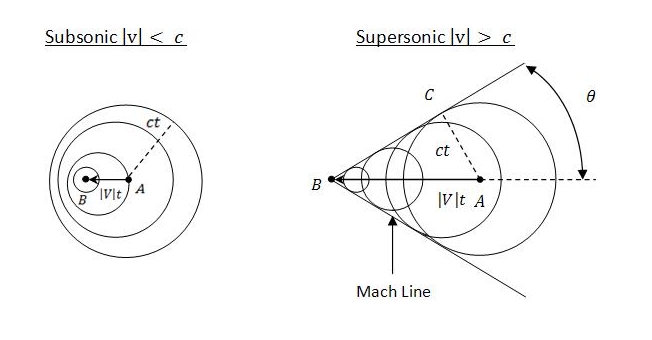}
 \caption{Wavefronts emitted in subsonic/subcritical regime (left) and formation of Mach cone in supersonic/supercritical regime (right).}
 \label{Fig:Mach-cone-theory}
\end{figure}

The question of whether and where the surface waves are blocked by the fluid flow could be answered by comparing the radial fluid velocity $v_s^r$ at the surface and the propagation velocity $c$ of the surface waves. However, given that we are not interested in the precise values of each quantity separately, but only in the ratio $v_s^r/c$, we have opted for the following alternative. Our method is based on the Mach cone well known in the case of sound waves, see Fig.~\ref{Fig:Mach-cone-theory}. There, an object propagating at a speed $V$ above the speed of sound $c_s$ leaves behind an observable cone, the Mach cone. The cone is formed by the envelope of the subsequent wavefronts emitted by the object, which partially escape from each other. The half-angle $\theta$ of the cone obeys $\sin \theta=ct/|V|t=c/|V|=1/M$, with $M$ the Mach number, see Fig.~\ref{Fig:Mach-cone-theory}. For an object propagating at a speed $V<c_s$, the subsequent wavefronts remain inside the previous ones and no Mach cone is formed. Note that the appearance of a Mach cone can be interpreted as a consequence of the Cerenkov threshold condition expressed as $\omega-{\bf k.V}=0$, or in other words $\omega-k.V\cos\psi=0$. Since the Cerenkov angle $\psi$ and the Mach angle $\theta$ are complementary, i.e. $\psi=\pi/2-\theta$, the same trigonometric rule $\sin \theta =c/|V|$ again follows immediately.

Exactly the same argument can be applied to the case of an object standing still at the surface of a fluid flow. By comparing the fluid flow velocity $v_s^r$ to the propagation speed of surface waves $c$, the following cases can occur:
\begin{itemize}
 \item Supercritical region: $v_s^r>c ~~\Rightarrow~~ \theta \in \left[0,\frac{\pi}{2}\right]$
 \item Subcritical region: $v_s^r<c ~~\Rightarrow~~ \theta$ complex; Mach cone disappears
 \item Horizon: $v_s^r=c ~~\Rightarrow~~ \theta=\frac{\pi}{2}$
\end{itemize}

\begin{figure}[!htpb]
 \centering
\includegraphics[width=.55\textwidth]{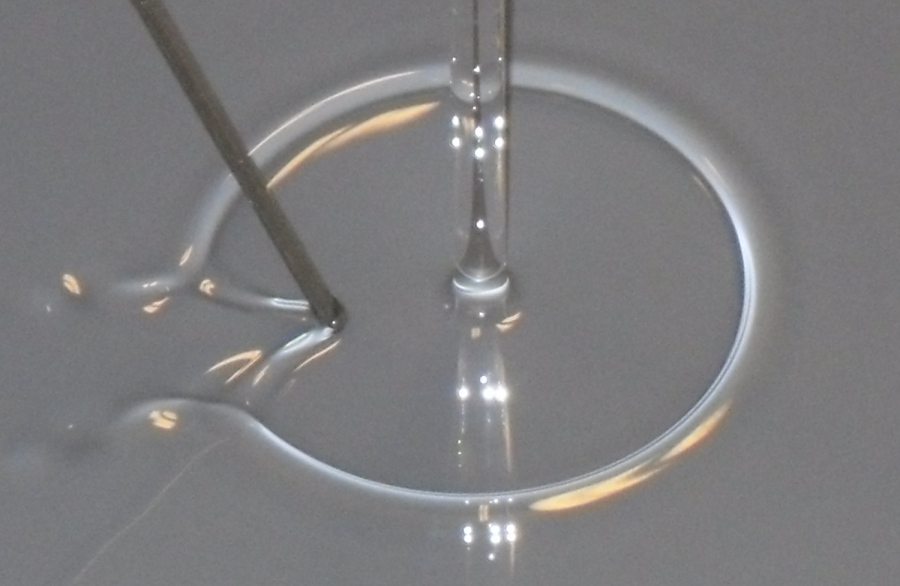}
 \caption{Circular hydraulic jump and Mach cone created by inserting a sharp object in the fluid flow.}
 \label{Fig:Mach-cone-experiment}
\end{figure}

\begin{figure}[!htpb]
 \centering
\includegraphics[width=.40\textwidth]{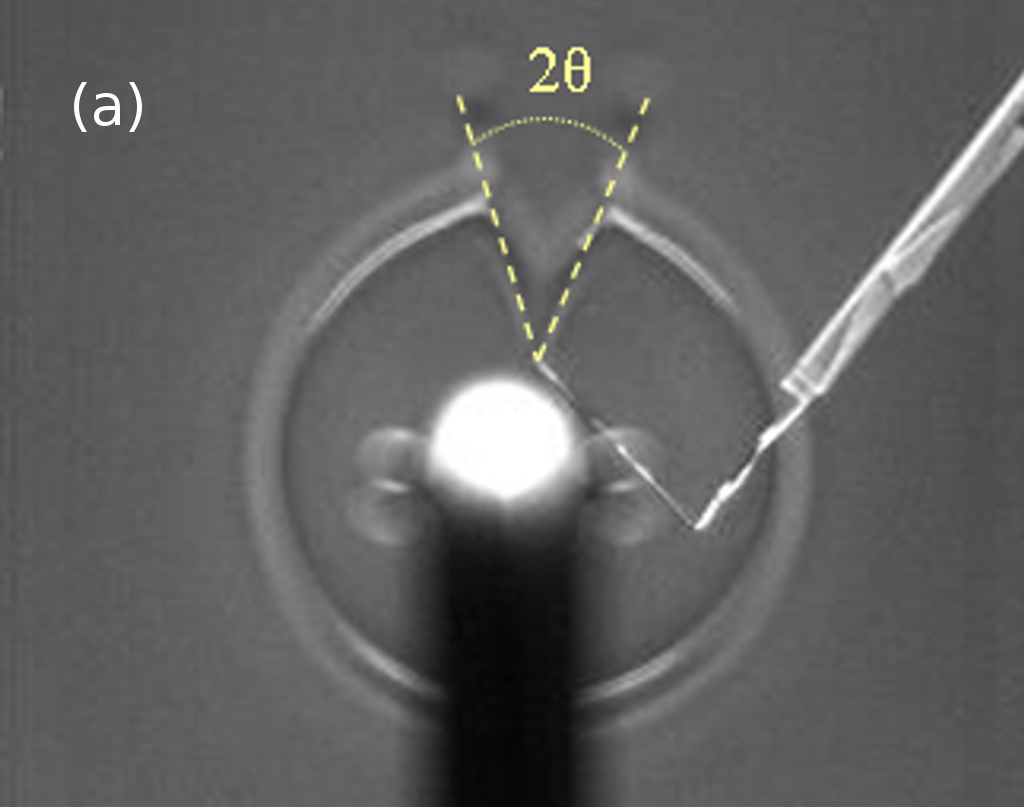}
\includegraphics[width=.40\textwidth]{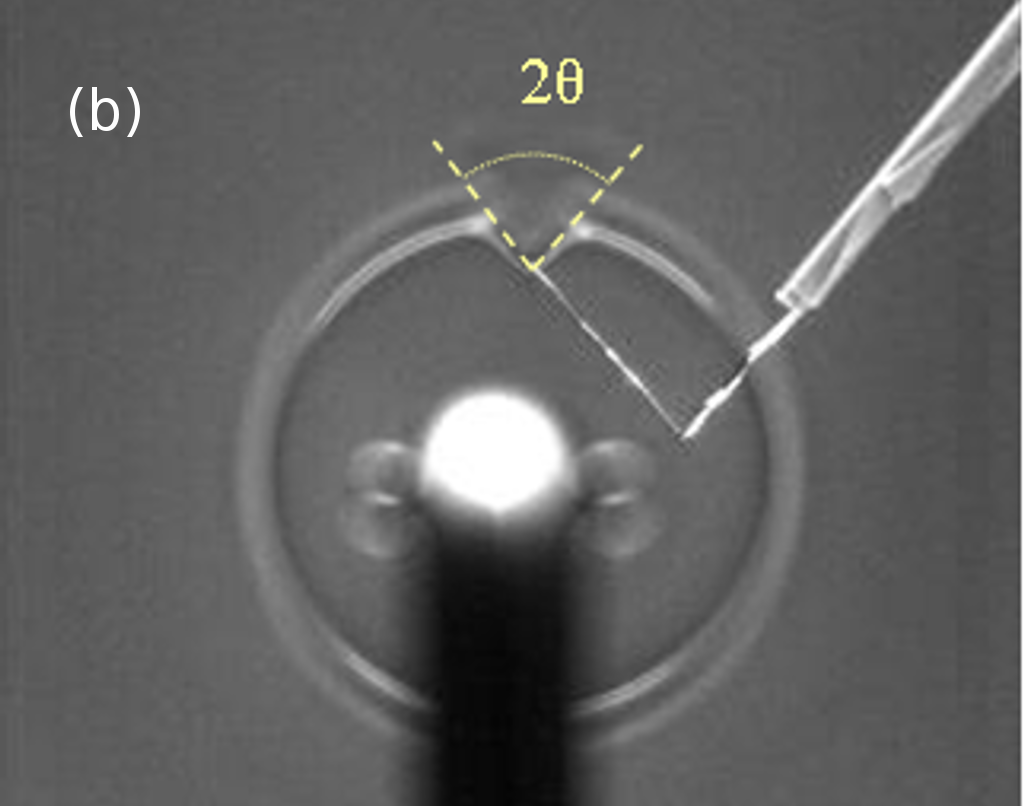}

\vspace*{0.5ex}
\includegraphics[width=.40\textwidth]{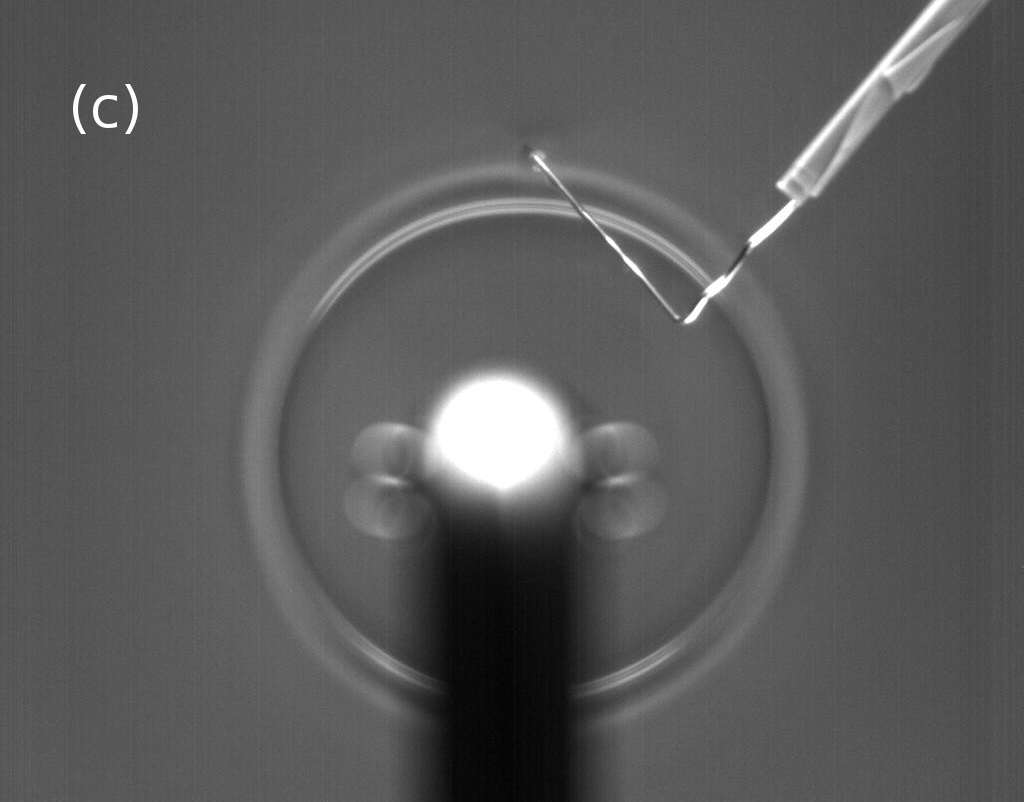}
\caption{
Measurements of the Mach angle $\theta$. A needle is placed inside the flow at varying distances from the centre of the jump. (a) Mach cone near the centre of the jump. (b) Mach cone near the edge of the jump. (c) The Mach cone disappears just outside the jump.}
\label{Fig:mach-cone2} 
\end{figure}

\begin{figure}[!htpb]
 \centering
\includegraphics[width=.65\textwidth]{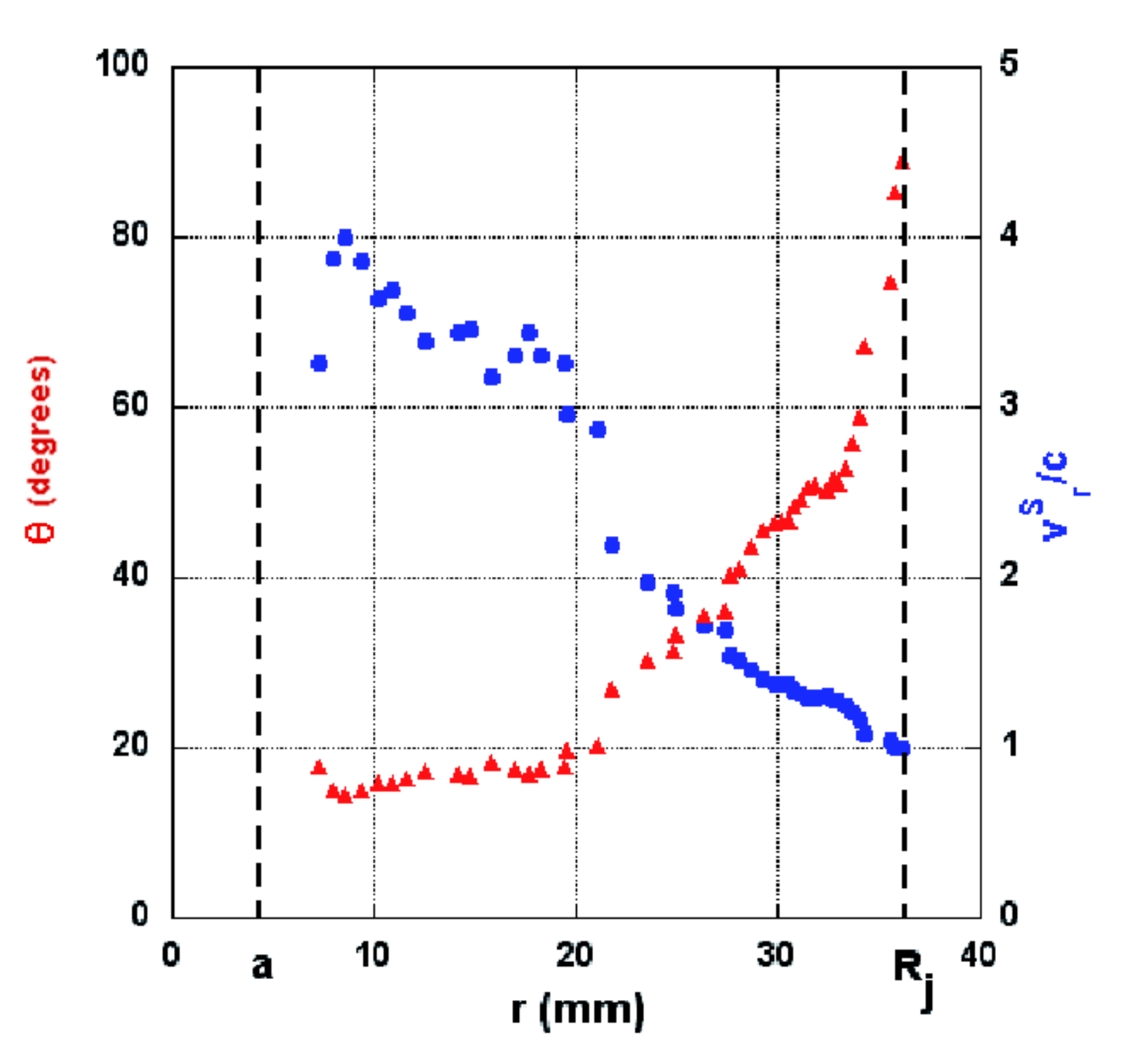}
\caption{
Mach angle $\theta$ (red triangles) and ratio $v_ r^s/c$ (blue circles) as a function of the distance $r$ from the centre of the jump. The dashed vertical lines represent the nozzle radius $a$ and the jump radius $R_j$.
Experimental parameters: see~\cite{Jannes:2010sa}.}\label{Fig:results-mach-angle} 
\end{figure}

Our experiment to demonstrate the presence of a hydrodynamic horizon is described in detail in~\cite{Jannes:2010sa}. Essentially, we have pumped silicon oil through a steel nozzle onto a horizontal PVC plate. A needle was placed such as to penetrate the flow surface at varying  distances $r$ from the centre of the circular jump, where the oil jet impacts on the PVC plate. For each position of the needle, we haved photographed the setup with a high-speed camera and measured the corresponding Mach angle $\theta$, see Fig.~\ref{Fig:mach-cone2}. The resulting angles and the derived ratio $v_ r^s/c$ are shown in Fig.~\ref{Fig:results-mach-angle}. Inside the jet impact zone ($r<a$, where $a$ is the nozzle radius, see also Fig.~\ref{Fig:jump-schema} above), we expect $v_r^s\ll c$ followed by a steep increase for $r\gtrsim a$ until a certain value $v_r^{s, \text{max}}$, since the fluid impacts vertically before being converted into a radial flow. The field of vision of our experimental setup starts near this maximum, see Fig.~\ref{Fig:results-mach-angle}, corresponding to a Mach angle $\theta$ of roughly $\pi/10$. From there, $\theta$ smoothly increases to about $\pi/4$ at approximately 3/4 of the jump radius, and then rapidly opens up to reach exactly $\pi/2$, and hence $v_r^s/c=1$, near the ridge of the jump. 


These results provide a clear proof that the circular hydraulic jump constitutes a two-dimensional hydrodynamic white hole: surface waves travelling at a velocity $c$ towards the jump from the exterior are trapped outside in precisely the same sense as light is trapped inside a gravitational black hole. The corresponding white-hole horizon is situated at the radius of the jump itself. Curiously though, in our experiments, the critical point $v_r^s/c=1$ is actually reached in a very smooth way. The ratio $v_r^s/c$ decreases mainly far from the jump, well inside the inner region. This is in striking contrast with the standard theoretical models in fluid mechanics, which describe the circular jump as a shock wave and therefore prescribe that the critical point itself should lie within a sharp (and in models without viscosity: discontinuous) transition from a supercritical to a subcritical regime, see e.g.~\cite{Bohr:1997}. It is not clear whether this smooth transition to the critical point is a genuine property of the jump itself, or a consequence of the perturbation of the flow pattern due to the insertion of the needle, which somehow smoothens out the shock wave (or shifts its position). But the main result stands clear: the circular jump is a white hole.

From the point of view of analogue gravity, the following features of this hydrodynamic white hole are particularly striking, especially in comparison with other current or planned experiments, e.g. in optics~\cite{Belgiorno:2010wn} or Bose-Einstein condensates~\cite{Lahav:2009wx}:
\begin{itemize}
 \item The white hole here is created ``spontaneously''. One only needs to arrange for a sufficient fluid flow rate, but no extraordinary engineering is required to obtain a long-lived horizon. In fact, it suffices in principle to open the tap in a kitchen sink to observe such a white hole. All other ingredients in the experimental setup (the choice of silicon oil, the precision of the pump etc) serve merely to make the experiment cleaner and free of perturbations and other undesired effects, but do not affect the essential point which is the spontaneous creation of a (hydrodynamic) white hole. This makes hydrodynamic white holes a rather exceptional example within the condensed-matter analogues, as in most analogue models there is a strong resistance against the creation of a horizon, see e.g.~\cite{Volovik:2003ga}, which is usually accompanied by the appearance of several types of instabilities which tend to destroy the horizon.
 \item The white hole can be observed with the naked eye: the location of the white hole horizon is precisely where the fluid undergoes the characteristic jump. This can be verified in the way described above, in principle simply by sticking a pencil in a kitchen sink. 
\end{itemize}
These two elements mean that the circular jump provides a nice example of a simple ``do-it-yourself black-hole experiment'' which can be explained in terms accessible to any person with a basic knowledge of trigonometry.

Let us now go back to the dispersion relation for surface waves and examine in which dispersive regime the circular jump typically lies.

\section{Dispersion relation}
\begin{figure}[!htpb]
 \centering
\includegraphics[width=.60\textwidth]{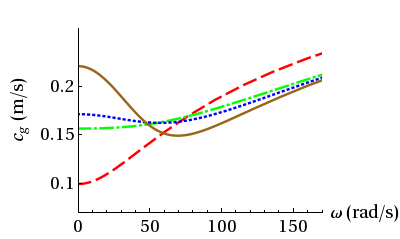}
\caption{\label{Fig:group-velocities} 
Group velocities (at $U=0$) for silicon oil (viscosity $\nu = 20 \text{cS} \approx 20 \nu_\text{water}$; surface tension $\gamma=0.0206\text{N/m}\approx\frac{1}{3}\gamma_\text{water}$; density $\rho =950 \text{kg}/\text{m}^3$). 
and different values of the fluid height $h$ [from bottom to top at low $\omega$]: $h=1$mm (red dashed line); 2.5mm (green dashed-dotted line); 3mm (blue dotted line); 5mm (full brown line). Typical fluid heights in the experiments reported in~\cite{Jannes:2010sa} are $\sim 1$mm, corresponding to the red dashed line and hence a purely superluminal correction to the dispersion relation.
}
\end{figure}

Developing Eq.~\eqref{disp-rel} for the case $kh\ll1$ and truncating at $\mathcal{O}(k^4)$, one obtains
\begin{align}\label{dispersion-full}
(\omega-Uk)^2\approx ~& ghk^2 + \left(\frac{\gamma h}{\rho}-\frac{gh^3}{3}\right) k^4+\mathcal{O}(k^6)
\\= ~& c^2k^2 + c^2\left(l_c^2-\frac{h^2}{3}\right)k^4+\mathcal{O}(k^6),
\end{align}
where $l_c=\sqrt{\gamma/g\rho}$ is the capillary length ($l_c=1.49$mm for the silicon oil in our experiments). The shallow-water regime is relativistic at low values of $k$ (the associated ``relativistic speed'' $c$ is $c=\sqrt{gh}$, i.e. the surface wave velocity in the low-$k$ or gravity limit where capillarity is negligible). For intermediate $k$'s, the dispersion can either be ``subluminal'' or ``superluminal'' (i.e., the group velocity $c_g\equiv \frac{d\omega}{dk}$ decreases or increases with $k$, respectively), see Fig.~\ref{Fig:group-velocities}, with a critical transition depth $h_\text{trans}= \sqrt{3}l_c$ (see also~\cite{Visser:2007du}), corresponding to an inflection point $d^2c_g/dk^2=0$ at $k=0$. At higher $k$'s (once $kh\sim 1$), the development~\eqref{dispersion-full} can no longer be trusted, and the dispersion becomes superluminal, irrespective of the value of $h$, asymptoting to $c_g \propto \sqrt{k}$. 

We are here interested in the intermediate regime, i.e. in the first non-relativistic corrections. 
For our silicon oil, $h_\text{trans}\approx 2.6$mm, while the typical heights of the fluid inside the jump are much smaller (typically $\sim 1$mm). The same relation $h<\sqrt{3}l_c$ holds true for experiments with other fluids reported in the literature (at least in the so-called ``type I'' circular jump, which avoids reversal currents and is therefore the most interesting from the point of view of analogue gravity, see~\cite{Jannes:2010sa}). This confirms that the circular jump is superluminal ($c_g$ increasing with $k$) for realistic experimental parameters. A direct observational test is the following. When introducing the needle in a circular jump inside a water flow, tiny capillary waves are observed {\it upstream} from the needle. In other words, these capillary waves manage to propagate some distance against the background flow, thereby confirming the superluminal character. Note that in water, $h_\text{trans}\approx 4.7$mm whereas typical fluid depths inside the jump are $\lesssim0.5$mm~\cite{Craik:1981}. In silicon oil, such capillary waves upstream from the needle are not observed, indicating that the superluminal correction is too small to compensate for the strong dissipation due to viscosity.

In such a superluminal regime, sufficiently high-frequency modes can penetrate across the horizon in the classically prohibited sense. The interior of the black/white hole is then no longer causally separated from the outside. Such superluminal horizon-crossing effects are considered in several scenarios for quantum gravity phenomenology. They are especially interesting in the sense that they put the robustness of several aspects of black hole physics (and in particular, Hawking radiation) to the test~\cite{Barcelo:2008qe}. Most of these are ultimately related to vacuum selection issues~\cite{Barcelo:2010bq,Barbado:2011dx}, which are absent in the case of the circular jump: There is a clear ``preferred'' reference frame, the laboratory reference frame, and hence a clear and well-defined vacuum state (the one in which the fluid stands still with respect to the laboratory). Nevertheless, some aspects of the robustness of the thermal spectrum of Hawking radiation with respect to trans-Planckian physics could in principle be tested in the circular hydraulic jump, see the discussion in~\cite{Jannes:2010sa}.

The main barrier to such studies is the fact that these penetrating modes are blueshifted by the countercurrent, and therefore strongly damped due to viscosity, making it extremely hard to measure them in a reliable way. Superfluids could provide an interesting way out of this problem. In fact, an earlier experiment with a circular jump in $^4$He~\cite{Rolley} has shown capillary ripples near the jump. Below the superfluid transition temperature, stationary capillary ripples also appeared well within the inner region. There also exists a concrete proposal for an experiment with ripplons at the interface between the superfluid $^3$He-A and $^3$He-B phases in a rotating cryostat~\cite{Volovik:2002ci}, related to an earlier experiment reported in~\cite{Blaauwgeers:2002}. The essential idea is to take advantage of the fact that the critical rotation velocity $\Omega_c$ for the appearance of vortices, which allow the superfluid component to co-rotate with the normal component, is lower for the A-phase than for the B-phase. A cylindrical sample with superfluid $^3$He is set into rotation, and separated into the A- and B-phase by the gradient of a magnetic field. A differential rotation $\bar{\Omega}$ between both superfluid phases is achieved at rotation velocities $\Omega_{c,A}<\Omega<\Omega_{c,B}$. By modulating the propagation velocity $c$ of the ripplons through a variation of the thickness of the B-layer channel, the interface flow velocity $v=r\bar{\Omega}$ can be made to locally exceed $c$ (typically of the order of 1 mm/s), thereby creating an ergoregion where $v>c$. For sufficiently thin B-channel layers, the relativistic limit corresponding to the shallow-water regime described earlier is recovered, and the boundaries $v=c$ of the ergoregion form a pair of relativistic black/white hole horizons. 

Such an experiment could be interesting not only to measure the radiation associated with these horizons, but also because it can be extended to simulate a black hole singularity. Indeed, at an even higher critical velocity, corresponding to the proper Kelvin-Helmholtz threshold~\cite{Volovik:2002ci}, the $g_{tt}$ component of the effective metric becomes infinite, indicative of an effective black hole singularity.

The question of dissipation is also related to the trans-Planckian issue, which we discuss next.

\section{Trans-Planckian problem}
In analogue gravity systems, the infinite blue-shifting associated to strictly relativistic horizons is avoided 
because of dispersive corrections at short wavelengths. One of the cornerstones of the analogue gravity programme has therefore from the start been that perhaps the same could occur in real gravity. 

Historically, the focus has been mainly on the study of subluminal (``normal'') dispersion. In that case, the trans-Planckian problem is avoided because the incident waves are reflected just before reaching the white horizon, since their group velocity decreases as they blueshift. For a black hole horizon, the argument should be time-reversed: late-time outgoing modes originate from finite-$k$ modes which start off outside the horizon and are redshifted as they are dragged towards the black hole. Due to this (finite) redshifting, their group velocity increases and they manage to escape away from the black hole.

Superluminal dispersion relations, on the other hand, solve the trans-Planckian problem by allowing blue-shifted waves to penetrate across the (low-$k$) horizon, rather than endlessly blueshifting against it, as in the purely relativistic case. However, if one imagines a flow with a counter-current which increases without bounds, then the superluminal waves will also be blueshifted without bounds as they propagate ever further across the horizon, against an ever faster counter-current. Again, for a black hole, the argument should be time-reversed, and outgoing superluminal modes seem to ultimately originate from the singularity.

An additional characteristic scale of the system is needed to avoid this scenario. In principle this could be some scale which puts a limit on the acceleration of the fluid. Once the counter-current reaches its maximum value, the superblueshifted capillary waves would then be allowed to freely propagate against the background flow. In practice, however, in the case of the circular jump in a classical liquid, the characteristic scale which avoids the endless blueshift is associated to dissipation: the blueshifted waves are strongly damped due to viscosity, typically very close to the horizon (just inside the jump). The trans-Planckian problem is thus solved through a combination of dispersion and dissipation. The analogue of the Planck scale (the intermolecular distance) is never reached and the fluid continuity approximation is never violated. 

It is an open question how the trans-Planckian problem is solved in real gravity, if at all. No clues of modified dispersion or other forms of Lorentz violation have been found so far, and there exist strong bounds on the possibility of Lorentz violations at scales at least up to $10^6E_\text{Planck}$, see e.g.~\cite{Jacobson:2005bg,Maccione:2009ju}. The question is then whether this means that trans-Planckian frequencies are really reached in the physics associated to black holes, implying that some theory of quantum gravity comes into play, or whether there exist other (non-dispersive) mechanisms to avoid such a trans-Planckian scenario. The latter was argued, for example, in \cite{Barcelo:transplanckian}. Assuming that black holes in the strict sense of general relativity (with long-lasting trapping horizons and a central singularity) do not form in astrophysics, but are replaced by ``black stars'', objects which hover just outside their own gravitational radius (maintained, for example, by vacuum polarization pressure~\cite{barcelo-fate}), it was argued that Hawking radiation could be produced by the internal pulsations of such objects without any appeal to trans-Planckian frequencies nor the need to invoke modifications of the dispersion relation.

\section{Outlook}
The proof that the circular jump constitutes a hydrodynamic white hole confirms Rayleigh's century-old conjecture that the circular jump marks the transition between a supercritical and a subcritical regime. The proof itself was based on well-known physics (the Mach cone) and used a straightforward methodology. This relative simplicity means that it can in principle be repeated in a kitchen sink. But this should not deceive anyone: further advance in the study of the circular jump as a hydrodynamic white hole is likely to be much more complicated and involve state-of-the-art technology. Detecting the radiation associated with this hydrodynamic horizon would require either precision-measurement of the blueshifted waves that appear inside the jump, or the use of viscous-free fluids such as liquid Helium near zero temperature. Whether such approaches will allow results similar to those obtained in the wave-tank experiments \cite{Rousseaux:2007is,Weinfurtner:2010nu} is an interesting question. The complication lies mainly in the dissipation of the blueshifted waves near the horizon. The circular jump has a double disadvantage. The first one is a matter of scales: it is easier to work  with long wavelengths, which are less prone to dissipation, in a linear water-tank than in a circular set-up. Second, and associated: the superluminal character of the dispersion relation in the case of the circular jump means that the Hawking process involves the physics inside the horizon, where there is additional blueshifting due to the accelerating counter-current, whereas in the subluminal regime of the water-tank, all the processes related to Hawking radiation take place outside the horizon. 

The question of whether it is possible at all to reproduce the hydrodynamic analogue of Hawking radiation in a situation which involves the physics inside the horizon, would also precisely make the circular jump an interesting test case. Several ``trans-Planckian'' issues involved and competing effects were discussed in \cite{Jannes:2010sa}. 

Whatever the outcome of such further work, it should be clear that the interest of cross-fertilization between fluid mechanics and gravity, contrarily to a relativistic horizon, is two-directional.

\section*{Acknowledgments}
The authors thank Romain Piquet, Philippe Ma\"issa and Christian Mathis for collaboration on the circular jump experiment. G.J. thanks INSMI, CNRS and the University of Nice, as well as a FECYT postdoctoral mobility contract of the Spanish MEC/MICINN, the Academy of Finland (Centers of Excellence Programme 2006-2011, grant 218211) and the EU 7th Framework Programme (FP7/2007-2013, grant 228464 Microkelvin). This research was supported by the R\'{e}gion PACA (Projet exploratoire HYDRO) and the Conseil G\'en\'eral 06.



\end{document}